\documentclass[fleqn,usenatbib]{mnras}

\usepackage{newtxtext,newtxmath}

\usepackage[T1]{fontenc}

\DeclareRobustCommand{\VAN}[3]{#2}
\let\VANthebibliography\thebibliography
\def\thebibliography{\DeclareRobustCommand{\VAN}[3]{##3}\VANthebibliography}

\usepackage{graphicx}	
\usepackage{amsmath}	

\title[Dust pressure in debris discs]{Why dust pressure matters in debris discs}

\author[E. M. Lynch, J. B. Lovell and A. A Sefilian]{
Elliot M. Lynch,$^{1}$\thanks{E-mail: elliot.lynch@ens-lyon.fr}
Joshua B. Lovell,$^{2}$
and Antranik A. Sefilian$^{3}$
\\
$^{1}$ Univ Lyon, Univ Lyon1, Ens de Lyon, CNRS, Centre de Recherche Astrophysique de Lyon UMR5574, F-69230, Saint-Genis,-Laval, France\\
$^{2}$ Centre for Astrophysics, Harvard \& Smithsonian, 60 Garden St, Cambridge, MA 02138, United States \\
$^3$Astrophysikalisches Institut und Universit{\"a}tssternwarte, Friedrich-Schiller-Universit\"at Jena, Schillerg\"a{\ss}chen~2--3, D-07745 Jena, Germany \\
}

\date{Accepted XXX. Received YYY; in original form ZZZ}

\pubyear{2022}

\begin{document}
\label{firstpage}
\pagerange{\pageref{firstpage}--\pageref{lastpage}}
\maketitle

\begin{abstract}
 There is a common assumption in the particulate disc community that the pressure in particulate discs is essentially zero and that the disc streamlines follow Keplerian orbits, in the absence of self-gravity or external perturbations. It is also often assumed that the fluid description of particulate discs is not valid in the presence of crossing orbits (e.g. from nonzero free eccentricities). These stem from the misconception that fluid pressure arises due to the (typically rare) collisions between particles and that the velocity of particles in fluids are single-valued in space. In reality, pressure is a statistical property of the particle distribution function which arises precisely because there is a distribution of velocities at a given position. In this letter we demonstrate, with simple examples, that pressure in particulate discs is non-zero and is related to the inclination and free eccentricity distributions of the constituent particles in the discs. This means many common models of debris discs implicitly assume a nonzero, and potentially quite significant, dust pressure. We shall also demonstrate that the bulk motion of the dust is not the same as the particle motion and that the presence of pressure gradients can lead to strong departures from Keplerian motion.
\end{abstract}




\section{Introduction}

Debris discs are circumstellar discs of dust and other solid particles typically left over from the dispersal of a gaseous protoplanetary disc, along with the collisional grind-down of planetesimal belts. Dynamically particles in the disc interact with the large bodies in the system through gravity and, in the case of stars, through the stellar radiation field and winds. While collisions between particles in the disc are important for setting the planetesimal and grain size distribution, they are typically not frequent enough to strongly affect the particle orbits \citep{Wyatt99}. Thus, the only important collective effect, on individual particle orbits, is that due to self-gravity \citep[see][]{Wyatt08,Matthews14,Hughes18}.

Observations, however, are not sensitive to individual particle orbits but are in fact sensitive to bulk (i.e. averaged) dust properties such as the dust surface density. The bulk motion of the dust can depart strongly from the motion of the individual dust particles as a result of the dust pressure gradients, which, as we shall show, can be important even in completely collisionless system. This can be particularly important in very narrow rings such as Fomalhaut, as imaged by \citet{Kalas05}, \citet{Boley12}, \citet{MacGregor17} and \citet{Gaspar23}; HD53143 \citep{MacGregor22,Stark23}; HR4796 \citep{Kennedy18,Schneider18,Olofsson19} and HD202628 \citep{Faramaz19}. The narrowness of the ring means that even a small dust pressure can result in order unity corrections to the bulk motion from the large pressure gradients.

In this letter we demonstrate the connection between dust pressure and the free eccentricity and inclination distributions in dust discs. We shall show how the presence of dust pressure can lead to the bulk dust motion strongly departing from Keplerian motion, even when every particle is constrained to follow a Keplerian ellipse. Finally, we shall discuss the importance of dust pressure in dynamical and morphological models of continuous debris discs \citep[e.g.][]{Davydenkova18,Marino19,Sefilian21,Sefilian23,Lovell21,Lovell22,Lynch22,Lovell23}.

The importance of pressure to the dynamics of completely collisionless systems is a well studied phenomena in galactic dynamics and has been known about since \citet{Jeans15}. The results we present here are a consequence of Jeans Equations and the dust pressure has a close relationship to the velocity ellipsoid used to characterise the distribution functions of galaxies \citep{Binney08}. In addition to Jeans equations, there is existing work, in the galactic dynamics literature, linking velocity dispersion to gravitational softening \citep{Miller71,Tremaine01,Touma02}. While the correspondence is not exact (velocity dispersion effects also occur in the absence of 2-body interactions), \citet{Tremaine01} has shown that velocity dispersion, gas pressure and potential softening have similar effects on spiral density waves propagating in particle discs.

In this letter we shall only consider dust pressure in the absence of gas. Dust pressure is also important in the presence of gas, particularly for larger grains or for time varying or turbulent flows. The presence of gas (particularly turbulent gas) makes the problem significantly more complicated and is unnecessary for understanding the main effects of dust pressure in gas-poor debris discs.

In Section \ref{inclination distribution} we shall show how dust pressure arises in a circular disc with an inclination distribution and how this leads to sub-Keplerian orbital motion. In Section \ref{free e distribution} we shall provide a brief sketch of how this effect works in a coplanar disc with nonzero free eccentricity. Finally Section \ref{discus} discusses our results in the context of existing models and Section \ref{conc} presents our conclusions.

\section{Inclination Distribution} \label{inclination distribution}

\begin{figure}
\centering
\includegraphics[trim=50 500 100 100, clip, width=\linewidth]{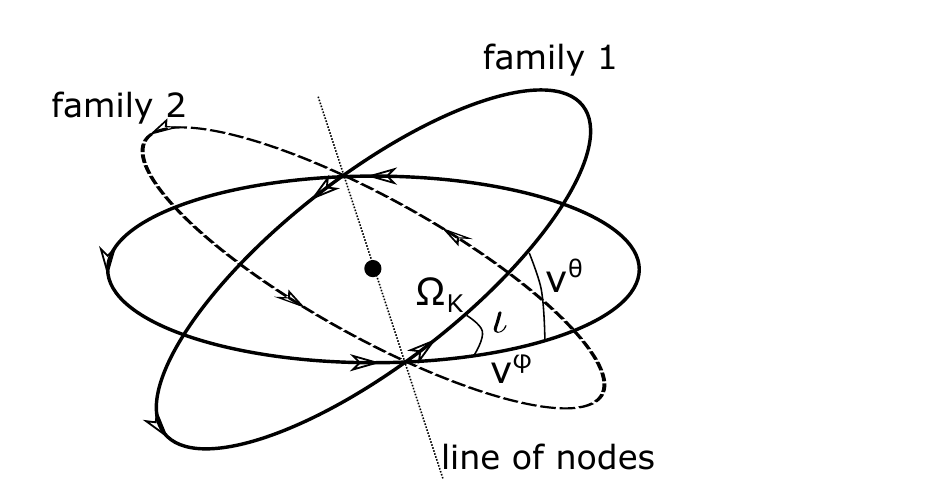}
 \caption{Two families of inclined orbits in a circular disc with constant inclination, $\iota$, and anti-aligned longitude of ascending nodes. At a given point on the midplane, if there is the same mass in family 1 and family 2, the averaged vertical velocity is zero and the average azimuthal velocity (as defined in Section \ref{inclination distribution}) is $u^{\phi} = \Omega_{K} \cos \iota$.}
 \label{Inclined orbit families}
\end{figure}

The simplest setup that can demonstrate the importance of dust pressure is to consider a disc where all the particles are on circular orbits with a constant inclination, $\iota$, with respect to the midplane and a random longitude of ascending node. Consider a point, $p$, on the midplane with (spherical) radius $r$, azimuthal angle $\phi$ and polar angle $\theta=\pi/2$. The particles which pass through this point will be split into two families, family 1 with the longitude of ascending node aligned with $\phi$ and family 2 where it is anti-aligned (see Figure \ref{Inclined orbit families}). Note that both of these families orbit in a prograde direction.

By simple geometric arguments, one can show that the velocity (in spherical polars) of particles in family 1, as they pass through p, is $\Omega \equiv v^{\phi} = \Omega_{K} \cos \iota$ and $v^{\theta} = \Omega_{K} \sin \iota$, where $\Omega_{K}$ is the circular Keplerian angular velocity at radius $r$. The velocity in family 2 is $v^{\phi} = \Omega_{K} \cos \iota$ and $v^{\theta} = -\Omega_{K} \sin \iota$, with both families having $v^{r} = 0$. Assuming the total mass of dust particles in family 1 \& 2 are the same then the bulk (i.e. mass weighted average) vertical velocity $u^{\theta} = 0$, and the bulk azimuthal velocity $u^{\phi} = \Omega_{K} \cos \iota$. Thus, the bulk motion of the dust in the midplane consists of circular sub-Keplerian motion, which only matches the particle motion when $\iota = 0$.

We now show how dust pressure gradients are responsible for the sub-Keplerian character of the velocity. As particles in family 1 all have the same orbital elements, the distribution function for these particles can be obtained as a product of delta distributions. The distribution for family 2 can be constructed similarly. Assuming the masses in family 1 and family 2 are the same, and the mass crossing the midplane on the spherical shell with radius, $r$, is $M_{r} (r)$ then the distribution function, $f$, for the particles on the midplane $(\theta = \pi/2)$ is

\begin{align}
\begin{split}
f = &\frac{1}{2} \Sigma^{\circ} (r) \delta (v^{r}) \delta (v^{\phi} - \Omega_{K} \cos \iota) \\
&\times \left[ \delta (v^{\theta} - \Omega_{K} \sin \iota) + \delta (v^{\theta} + \Omega_{K} \sin \iota)\right] , 
\end{split} \label{distribution function}
\end{align}
where $\Sigma^{\circ} = \frac{M_{r}}{2 \pi r}$ and $\delta(x)$ is the Dirac-delta distribution. One can take velocity moments of Equation \ref{distribution function} to calculate the bulk properties of the particulate disc (which is typically what observations are sensitive to). The zeroth moment $\Sigma = \int f \, d v^{r} \, d v^{\phi} \, d v^{\theta} = \Sigma^{\circ}$ is the surface density on the midplane. The bulk velocity components, $u^{i}$, are given by $u^{i} = \frac{1}{\Sigma}  \int v^{i} f \, d v^{r} \, d v^{\phi} \, d v^{\theta}$, with $i=\{r,\phi,\theta\}$, which, as expected, yield the same velocities described in the previous paragraph.

We can also compute the second velocity moment of the distribution function, the dust pressure tensor, $P^{i j}$, which arises irrespective of the collisionality of the particles,

\begin{equation}
 P^{i j} = \int (v^{i} - u^{i}) (v^{j} - u^{j})  f \, d v^{r} \, d v^{\phi} \, d v^{\theta} . \label{pressure tensor def}
\end{equation}
As all particles have $v^{r} = u^{r} = 0$ and $v^{\phi} = u^{\phi}$, the components of the pressure tensor which involve $r$ or $\phi$ are zero and the only nonzero pressure tensor component is $P^{\theta \theta}$ as $u^{\theta} = 0$. However, the individual orbits have velocities of $v^{\theta} = \pm \Omega_{K} \sin \iota$, which results in the following pressure tensor component,

\begin{align}
\begin{split}
 P^{\theta \theta} &= \int (v^{\theta})^2  f \, d v^{r} \, d v^{\phi} \, d v^{\theta} , \\
&= \Sigma^{\circ} \Omega_K^2 \sin^2 \iota .
\end{split} \label{pthetatheta eq}
\end{align}
The momentum equation for the bulk motion of the dust, which arises from the first velocity moment of the (collisionless) Boltzmann equation, is

\begin{equation}
\Sigma D \mathbf{u} = -\Sigma \nabla \Phi - (\nabla \cdot \mathbf{P}) , \label{momentum equation inc}
\end{equation}
where $\Phi$ is the gravitational potential and $D = \partial_t + \mathbf{u} \cdot \nabla$ is the Lagrangian time derivative with respect to the dust flow. In galactic dynamics, Equation \eqref{momentum equation inc} is known as Jeans equation \citep{Jeans15,Binney08}. The last term, $- (\nabla \cdot \mathbf{P})$, is the divergence of the pressure tensor.  In a gas disc, this corresponds to the usual pressure gradient term $- \nabla P$. In dust discs (e.g. gas-poor debris discs) this term is usually neglected with the justification that the only force on the particle is due to the point mass potential of the central star. However, such a justification is incorrect as $- (\nabla \cdot \mathbf{P})$ is not due to the forces on the particles but is instead a statistical property that arises as a consequence of the particle velocities being multivalued at any given position in the disc. For the pressure tensor associated with our distribution function (Equation \ref{distribution function}) this is 

\begin{equation}
 (\nabla \cdot \mathbf{P}) = -r P^{\theta \theta} \hat{\mathbf{e}}_r,
\end{equation}
close to the midplane, where $\hat{\mathbf{e}}_r$ is the radial unit vector. Looking for the steady solutions to the momentum equation (Equation \ref{momentum equation inc}), near the midplane, one finds:

\begin{align}
-r [ \Omega^2 + (u^{\theta})^2] &= -r \Omega_K^2 + r \frac{P^{\theta \theta}}{\Sigma^{\circ}} , \label{r equation} \\
r^{-1} u^{r} \Omega &= 0 , \label{phi equation} \\
r^{-1} u^{r} u^{\theta} &= 0 .
\end{align}
By Equation \eqref{phi equation}, $u^{r} = 0$ and by symmetry we require $u^{\theta} = 0$. Substituting the expression for $P^{\theta \theta}$ into Equation \eqref{r equation} we arrive at $\Omega^2 = \Omega_K^2 \cos^2 \iota$, which matches that derived directly from the distribution function. Thus, we see that the pressure tensor gradients are necessary to obtain the correct velocity for the bulk motion, which is sub-Keplerian due to pressure support, similar to gas discs. If one were to instead neglect the pressure gradients in the momentum equation, one would instead obtain $\Omega=\Omega_K$ and incorrectly arrive at a Keplerian bulk dust motion.

This result will carry over to a more realistic inclination distribution \citep[e.g.][]{Brown01,Matra19} with the added complication of evaluating the integral in Equation \ref{pthetatheta eq}. Finally, in discs where only one of the two families exist, there are no intersections between particle orbits and the velocity remains single valued. Thus, the pressure term vanishes, $\mathbf{P} = 0$, and, by Equation \eqref{phi equation} one has $\Omega^2 + (u^{\theta})^2 = \Omega_K^2$ corresponding to an inclined Keplerian disc.

\section{Free Eccentricity Distribution} \label{free e distribution}

\begin{figure}
\centering
\includegraphics[trim=50 520 100 20, clip, width=\linewidth]{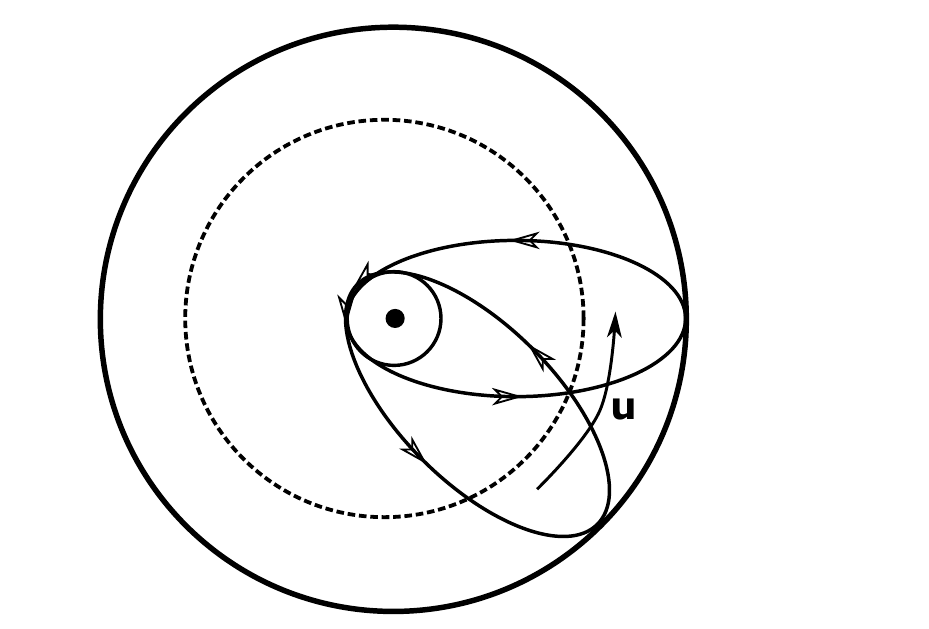}
 \caption{In a disc with only free eccentricity, the bulk motion of the dust consists of circular, non-Keplerian motion.}
 \label{free ecc cartoon}
\end{figure}

Another situation where dust pressure is important is in axisymmetric discs with nonzero free eccentricity (see Figure \ref{free ecc cartoon}). For simplicity, one can consider the limiting case where the semimajor axis distribution is infinitesimally narrow. Here, the radial velocity of the particle orbits averages to zero. Thus, the bulk fluid motion consists of circular, non-Keplerian, motion. That the bulk dust motion is non-Keplerian should be obvious from the fact that the inner disc edge is located at the pericentre of the particle orbits and, therefore, the particle velocities must be higher than the circular Keplerian values (similarly, the orbits are slower than the circular Keplerian value on the outermost orbit).

\begin{figure}
\centering
\includegraphics[trim=0 0 0 0, clip, width=\linewidth]{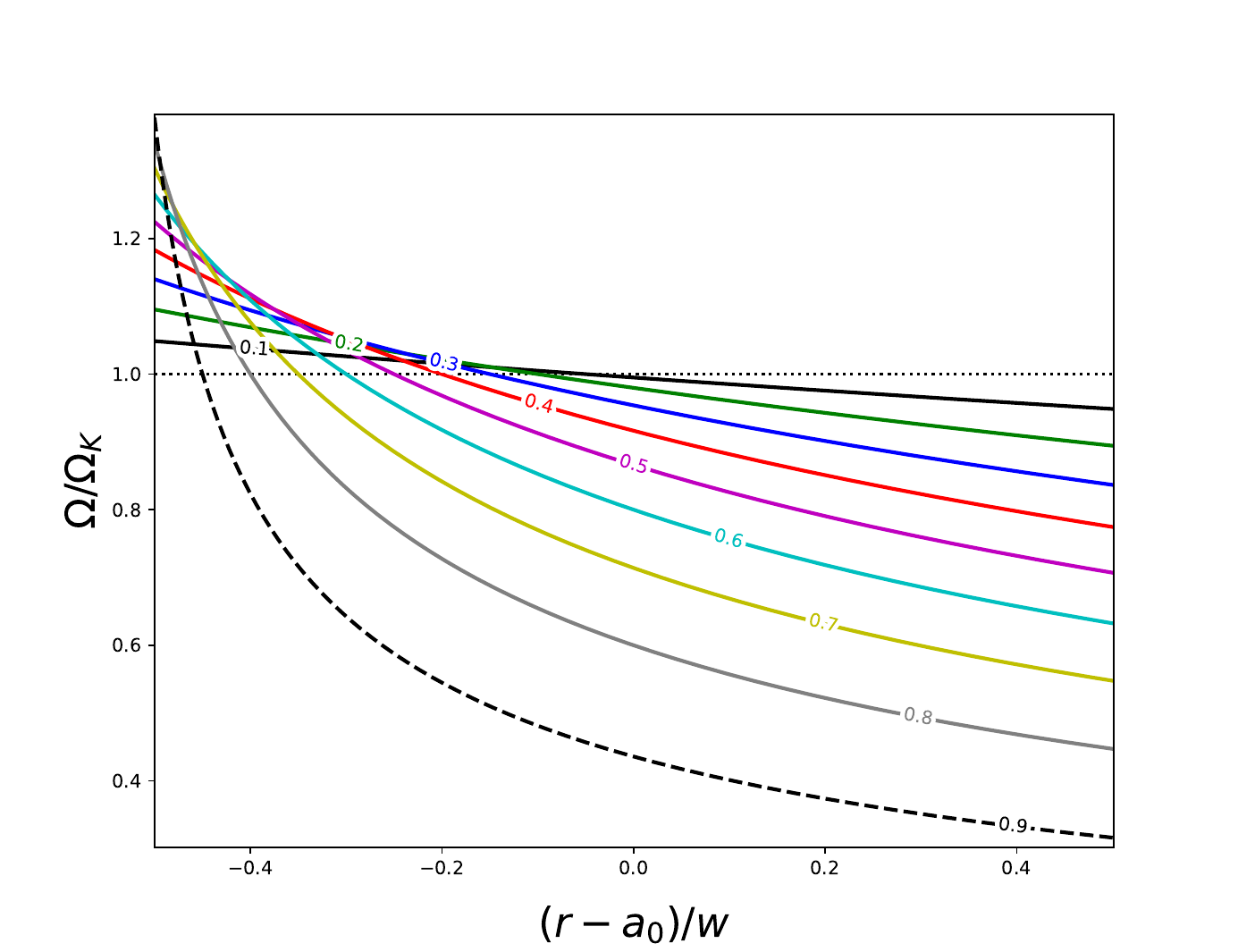}
 \caption{Angular velocity in a disc with free eccentricity from $e_p = 0.1$ (solid black line) to $e_p = 0.9$ (dashed black line) in intervals of $0.1$. $a_0$ is the, arbitrary, semimajor axis of the orbits and $w = 2 a_0 e_p$ is the ring width. The discs are super-Keplerian at small radii and sub-Keplerian at large radii. For the most eccentric discs the angular velocity is limited to being less than twice the Keplerian value in order for the particles to remain bound, however, the outer regions of the disc can become almost entirely pressure supported with an arbitrarily low rotation rate. In galactic dynamics this effect is a well known consequence of the virial theorem \citep{Binney08}.}
 \label{angular v free ecc}
\end{figure}

As with the inclination distribution, if one considers a point, $p$, with cylindrical radius, $r$, and azimuthal angle, $\phi$, then there are two orbits which pass through this point (as depicted Figure \ref{free ecc cartoon}); one with a velocity of $v^{r} =\sqrt{\frac{G M}{\lambda}} e \sin \psi (r,\phi)$ and $v^{\phi} = \sqrt{\frac{G M}{\lambda^3}}  [1 + e \cos \psi (r,\phi)]^2$; and the second with the same $v^{\phi}$ but $v^{r} =-\sqrt{\frac{G M}{\lambda}} e \sin \psi (r,\phi)$ (where $\lambda$ and $\psi (r,\phi)$ are the semi-latus rectum and true anomaly of the particle, respectively). This results in the (mass-weighted) mean velocity of $u^{r} = 0$ and $u^{\phi} = \Omega_{K} \sqrt{\frac{\lambda}{r}}$. Figure \ref{angular v free ecc} shows how this angular velocity compares with the Keplerian value for different eccentricities. It is evident that the angular velocity is super-Keplerian in the inner region of the disc and sub-Keplerian (potentially highly so) in the outer region.

We shall not repeat the same calculation we did with the circular disc with an inclination distribution (i.e. Equation \ref{pressure tensor def}), as the integration to obtain the pressure tensor is more involved. This is because, even if we consider the simplest case of a set of a disc with a constant free eccentricity and an infinitesimally narrow semimajor axis distribution \citep[so that each individual ring has a constant surface density,][]{Statler99}, these rings cannot be directly summed in phase space in the orbital coordinate system, as the orbital coordinates are functions of both position and velocity. Regardless of the difficulty in analytically evaluating the pressure tensor, the bulk velocity within the disc will satisfy the momentum equation (Equation \ref{momentum equation inc}), provided the pressure gradient term is included.

\section{Validity of morphological and dynamical models} \label{discus}

\subsection{Implication for particle debris disc models}

Debris disc observations are often interpreted using Monte-Carlo particle based-methods, which sample the orbits in the disc and typically include constant free and forced eccentricities \citep[e.g.][]{MacGregor17,Kennedy20,MacGregor22}. By sampling the particle distribution directly, this method implicitly accounts for the dust pressure. This will, however, mean the bulk motion around the ring is non-Keplerian, particularly when the ring is narrow due to the large pressure gradients present (for instance, the bulk motion in the disc models of Fomalhaut and HD53143 by \citet{MacGregor17} and \citet{MacGregor22}, respectively, have strong departures from Keplerian motion). One challenge for Monte-Carlo models is their slow convergence at high resolution \citep{Rafikov23,Lovell23}. This, however, will be further compounded if the velocity distribution needs to be adequately sampled to account for the effects of dust pressure.

\subsection{Implication for parameterised debris disc models} \label{param models}

One question that the fluid description may help to answer is the question of the validity of simplified models for debris disc morphology in the presence of orbital intersections, such as those developed by \citet{Marino19} and \citet{Lynch22} (and pioneered by \citet{Tremaine95,Statler99a,Statler99,Peiris03} in the galactic dynamics context). The model of \citet{Lynch22} postulates that the density in the debris disc satisfies the continuity equation for a fluid flow which follows a set of non-overlapping, confocal, elliptical Keplerian orbits - similar to the eccentric gas disc theories of \citet{Ogilvie01}, \citet{Ogilvie14} and \citet{Ogilvie19}. The validity of this assumption is thus likely the same as that for a fluid disc, i.e. they should provide a good description of the bulk dust properties on an orbital timescale when the forces due to pressure gradients are weak enough, relative to the Keplerian terms, to act on secular timescales. This is likely to be the case for geometrically thin (scale height $H \ll R$ the radial lengthscale), but not for radially wide debris discs (with radial lengthscales much longer than $H$) with small to moderate free eccentricities. As such, the model presented in \citet{Lynch22} will break down for geometrically thick, or sufficiently narrow debris discs (i.e. discs with $H \sim R$ or radial variations on lengthscales comparable with $H$), in the same way that the secular theories for eccentric gas discs break down under similar circumstances \citep{Ogilvie01,Ogilvie14,Ogilvie19}. 

The model of \citet{Pan16} models eccentric discs using 1D line densities, and predicts the disc surface brightness at long wavelengths to be $\propto T_{\rm eff}/v_{\rm orb}$, where $T_{\rm eff}$ is equilibrium temperature of the disc and $v_{\rm orb}$ is the orbital velocity. This is on a less secure footing as the dust does not obey the zero-th moment equation. \citet{Lynch22} showed that the model of \citet{Pan16} gives the correct behaviour in the low (observational) resolution limit of the more complete theory. Thus, the requirements for the \citet{Pan16} model to be valid are very restrictive in that the dust pressure must be low, whilst the beam size must be much larger than the ring width, but not so large that it is comparable with the disc radius. The issue with applying the model of \citet{Pan16} to a very narrow disc is that a relatively small free eccentricity, or inclination distribution, will lead to non-Keplerian bulk dust motion around the ring. We note that in situations where the model of \citet{Pan16} is valid, so too must be the models of \citet{Lynch22}. Finally, we note that \citet{Lovell23} recently presented a parameterised model which includes free eccentricity and thus implicitly includes dust pressure, similar to the particle based models.

\subsection{Implication for dynamical debris disc models}

For dynamical models of debris discs \citep[e.g.][]{Wyatt99,Davydenkova18,Sende19,Sefilian21,Sefilian23}, the presence of dust pressure is potentially even more important. The morphological models, discussed in Section \ref{param models}, are primarily interested in describing dust properties on orbital timescales, in order to obtain the dust surface density, and thus require very large pressure gradients to be strongly modified by the dust pressure. Dynamical models, however, are often concerned with describing the evolution of the disc on secular timescales. This means pressure can be much weaker and still significantly affect the dynamics on the timescales of interest. Notably, for lower mass discs, it is likely that pressure, not self-gravity, is the dominant collective effect and may need to be accounted for in models like those in \citet{Davydenkova18}, \citet{Sefilian21} and \citet{Sefilian23}. In fact, a simple consideration of the virial balance implies that the correction to the (bulk) angular velocity for a typical disc inclination distribution can greatly exceed that due to self-gravity.

To demonstrate the above, we consider a pessimistic scenario (a thin but massive disc), where the disc mass is $M_d = 3 \times 10^{-3} M_{*}$ (this includes the mass of the planetesimals which contain the large majority of the total disc mass) and a mean particle inclination above/below the midplane of $\iota \approx H/R = 0.02$. We can then estimate the angular velocity correction due to self gravity to be

\begin{align}
\begin{split}
 |\omega_{\rm SG}/\Omega_{K}|  &\approx \left( \frac{G M_d}{R^3}\right)^{1/2}  \left( \frac{G M_{*}}{R^3}\right)^{-1/2} \\
&= \sqrt{\frac{M_{\rm d}}{M_{\rm *}}} \lesssim 5 \times 10^{-2} \left( \frac{M_{\rm d}}{3 \times 10^{-3} M_{*}}\right)^{1/2} .
\end{split}
\end{align}
Similarly, based on the results from Section \ref{inclination distribution}, the angular velocity correction due to dust pressure is approximately

\begin{equation}
 |\omega_{\rm p}/\Omega_{K}|  \approx \sqrt{1 - \cos^2 \iota} \approx \iota \gtrsim 2 \times 10^{-2}\left( \frac{H/R}{0.02} \right) ,
\end{equation}
where we have neglected any contribution from free eccentricity. Thus we see that, even in this pessimistic scenario, the angular velocity correction from dust pressure is comparable to that from self-gravity. For thicker or less massive discs the correction due to pressure can potentially exceed that due to self-gravity by several orders of magnitude (for similar reasons, the effects of dust pressure can also exceed that due to some planetary perturbations). There is, however, a crucial difference between dust pressure and self-gravity in that self gravity is able to change the orbits of the particles, whereas pressure in the collisionless/non-interacting limit is not. It is primarily through its interaction with other dynamical processes that the influence of pressure on the dynamics of the disc will be felt. The dust pressure is likely the physical mechanism by which the disc self-gravity is softened (where softened gravity is required) when the finite size of the particles is unimportant (\citet{Tremaine01}; see also \citet{Sefilian19} for a detailed discussion of various softening prescriptions employed in nearly Keplerian discs. 

A dynamical model where dust pressure is implicitly present is \citet{Wyatt99}. \citet{Wyatt99} has shown that the (linear) perturbations from an eccentric planet result in an eccentric debris disc where the complex eccentricity of the particles is a sum of free and forced eccentricity. In the fluid picture, the planet produces a force in the momentum equation which affects the dust bulk velocity (resulting in a forced eccentricity), along with heating the dust fluid. The fluid thermal motion takes the form of the distribution of free eccentricities in the disc. There is no distinction between dynamical heating and thermodynamic heating in the large particle number limit. Softened N-ring models \citep{Touma02,Hahn03,Sefilian23} also imply the presence of the dust pressure, but only when the rings overlap (although the softening is motivated by velocity dispersion \`{a} la \citet{Tremaine01} and \citet{Hahn03}). This results in a subtle difference with \citet{Wyatt99} in that free eccentricity in a single ring in \citet{Wyatt99} results in an axisymmetric disc with dust pressure (similar to that considered in Section \ref{free e distribution}), while in \citet{Sefilian23} one instead has a zero pressure precessing eccentric ring. The dynamics of a (softened) N-ring model when the rings share a common semimajor axis has been considered by \citet{Touma02,Touma12,Touma14}. Notably \citet{Touma14} showed that, when self gravity is included, the thermal axisymmetric eccentricity distribution (as considered in Section \ref{free e distribution}) becomes unstable to developing an asymmetry, with azimuthal wavenumber $m=1$, if the particle eccentricities are low enough.

Finally, one consequence of the nonzero dust pressure in debris discs is the existence of dust sound/density waves. These sound waves are closely linked to the possibility of having an arbitrary initial mass distribution around an orbit. In relatively wide debris discs, free sound waves quickly become extremely tightly wound as a result of phase mixing on $\sim10$'s of orbits and have little effect on observations or dynamics. However, if the disc's spread in semimajor axes is small (making phase mixing slow), or the waves are being actively excited (e.g. by a planet), these waves may have important dynamical and observational effects that are worth exploring further. It is plausible that the spiral feature seen in the simulation of \citet{Sende19} can be interpreted as a form of spiral-density wave, excited by the planet, but propagating in the disc as a consequence of dust pressure, rather than the more usually considered self-gravity (with the added complication of radiation pressure affecting the orbits of small particles and artificial diffusion in orbital element space required for stability).

In conclusion, it should be possible to rederive several well-known results of debris disc dynamics from the fluid perspective (e.g. the shaping of an eccentric ring by a companion as described in \citet{Wyatt99}). Of course, unlike fluid discs, there is no guarantee that one can find closed form expressions for the evolution of the dust pressure\footnote{There does exists an exact closure for the Keplerian test particle problem when the collisionless Boltzmann equation is written in orbital elements space which might prove useful for this.}, so it remains to be seen whether there are aspects of debris disc dynamics which are easier to treat with the fluid description. A closure, when two body relaxation is included was obtained by \citet{Touma14}. We leave the task of finding a closed form fluid description of debris discs for future work. We note that one can sidestep the closure problem by directly working with the distribution function, as the fluid description is just a tool for interpreting and/or solving the collisionless Boltzmann equation.

\vspace{-2em} 

\section{Conclusion} \label{conc}

We have shown, with two simple examples, the importance of dust pressure in dust discs, even when the individual particles are on exact Keplerian orbits. Our findings have important implications for the study of gas-poor debris discs. The pressure in debris discs is not related to additional forces on particles, or collisions between particles, but is instead related to the non-zero velocity dispersion present in debris discs. One can interpret the inclination and free eccentricity distributions expected in typical debris discs to be equivalent to (anisotropic) dust pressure.  The presence of dust pressure in debris discs means that the bulk dust motion does not match the particle orbits and is not, generally, Keplerian. Finally, observations are primarily sensitive to bulk dust properties not individual particle orbits, so dust pressure is likely important in their interpretation.

\vspace{-2em} 

\section*{Acknowledgements}

We thank the reviewer for their suggestions to improve the manuscript clarity. This research was supported by the ERC through the CoG project PODCAST No 864965. This project has received funding from the European Union’s Horizon 2020 research and innovation program under the Marie Skłodowska-Curie grant agreement No 823823. J. B. Lovell would like to thank the Smithsonian Institute for a Submilimeter Array Fellowship. A. A. Sefilian acknowledges support by the Alexander von Humboldt Foundation through a Humboldt Research Fellowship for postdoctoral researchers.

\vspace{-2em} 

\section*{Data Availability}

No new data were generated or analysed in support of this research.

\vspace{-2em} 



\bibliographystyle{mnras}
\bibliography{dust_pressure_in_debris_discs} 

\begin{thebibliography}{}
\makeatletter
\relax
\def\mn@urlcharsother{\let\do\@makeother \do\$\do\&\do\#\do\^\do\_\do\%\do\~}
\def\mn@doi{\begingroup\mn@urlcharsother \@ifnextchar [ {\mn@doi@}
  {\mn@doi@[]}}
\def\mn@doi@[#1]#2{\def\@tempa{#1}\ifx\@tempa\@empty \href
  {http://dx.doi.org/#2} {doi:#2}\else \href {http://dx.doi.org/#2} {#1}\fi
  \endgroup}
\def\mn@eprint#1#2{\mn@eprint@#1:#2::\@nil}
\def\mn@eprint@arXiv#1{\href {http://arxiv.org/abs/#1} {{\tt arXiv:#1}}}
\def\mn@eprint@dblp#1{\href {http://dblp.uni-trier.de/rec/bibtex/#1.xml}
  {dblp:#1}}
\def\mn@eprint@#1:#2:#3:#4\@nil{\def\@tempa {#1}\def\@tempb {#2}\def\@tempc
  {#3}\ifx \@tempc \@empty \let \@tempc \@tempb \let \@tempb \@tempa \fi \ifx
  \@tempb \@empty \def\@tempb {arXiv}\fi \@ifundefined
  {mn@eprint@\@tempb}{\@tempb:\@tempc}{\expandafter \expandafter \csname
  mn@eprint@\@tempb\endcsname \expandafter{\@tempc}}}

\bibitem[\protect\citeauthoryear{{Binney} \& {Tremaine}}{{Binney} \&
  {Tremaine}}{2008}]{Binney08}
{Binney} J.,  {Tremaine} S.,  2008, {Galactic Dynamics: Second Edition}

\bibitem[\protect\citeauthoryear{{Boley}, {Payne}, {Corder}, {Dent}, {Ford}  \&
  {Shabram}}{{Boley} et~al.}{2012}]{Boley12}
{Boley} A.~C.,  {Payne} M.~J.,  {Corder} S.,  {Dent} W.~R.~F.,  {Ford} E.~B.,
  {Shabram} M.,  2012, \mn@doi [\apjl] {10.1088/2041-8205/750/1/L21}, \href
  {https://ui.adsabs.harvard.edu/abs/2012ApJ...750L..21B} {750, L21}

\bibitem[\protect\citeauthoryear{{Brown}}{{Brown}}{2001}]{Brown01}
{Brown} M.~E.,  2001, \mn@doi [\aj] {10.1086/320391}, \href
  {https://ui.adsabs.harvard.edu/abs/2001AJ....121.2804B} {121, 2804}

\bibitem[\protect\citeauthoryear{{Davydenkova} \& {Rafikov}}{{Davydenkova} \&
  {Rafikov}}{2018}]{Davydenkova18}
{Davydenkova} I.,  {Rafikov} R.~R.,  2018, \mn@doi [\apj]
  {10.3847/1538-4357/aad3ba}, \href
  {https://ui.adsabs.harvard.edu/abs/2018ApJ...864...74D} {864, 74}

\bibitem[\protect\citeauthoryear{{Faramaz} et~al.,}{{Faramaz}
  et~al.}{2019}]{Faramaz19}
{Faramaz} V.,  et~al., 2019, \mn@doi [\aj] {10.3847/1538-3881/ab3ec1}, \href
  {https://ui.adsabs.harvard.edu/abs/2019AJ....158..162F} {158, 162}

\bibitem[\protect\citeauthoryear{{G{\'a}sp{\'a}r} et~al.,}{{G{\'a}sp{\'a}r}
  et~al.}{2023}]{Gaspar23}
{G{\'a}sp{\'a}r} A.,  et~al., 2023, \mn@doi [Nature Astronomy]
  {10.1038/s41550-023-01962-6}, \href
  {https://ui.adsabs.harvard.edu/abs/2023NatAs...7..790G} {7, 790}

\bibitem[\protect\citeauthoryear{{Hahn}}{{Hahn}}{2003}]{Hahn03}
{Hahn} J.~M.,  2003, \mn@doi [\apj] {10.1086/377195}, \href
  {https://ui.adsabs.harvard.edu/abs/2003ApJ...595..531H} {595, 531}

\bibitem[\protect\citeauthoryear{{Hughes}, {Duch{\^e}ne}  \&
  {Matthews}}{{Hughes} et~al.}{2018}]{Hughes18}
{Hughes} A.~M.,  {Duch{\^e}ne} G.,   {Matthews} B.~C.,  2018, \mn@doi [\araa]
  {10.1146/annurev-astro-081817-052035}, \href
  {https://ui.adsabs.harvard.edu/abs/2018ARA&A..56..541H} {56, 541}

\bibitem[\protect\citeauthoryear{{Jeans}}{{Jeans}}{1915}]{Jeans15}
{Jeans} J.~H.,  1915, \mn@doi [\mnras] {10.1093/mnras/76.2.70}, \href
  {https://ui.adsabs.harvard.edu/abs/1915MNRAS..76...70J} {76, 70}

\bibitem[\protect\citeauthoryear{{Kalas}, {Graham}  \& {Clampin}}{{Kalas}
  et~al.}{2005}]{Kalas05}
{Kalas} P.,  {Graham} J.~R.,   {Clampin} M.,  2005, \mn@doi [\nat]
  {10.1038/nature03601}, \href
  {https://ui.adsabs.harvard.edu/abs/2005Natur.435.1067K} {435, 1067}

\bibitem[\protect\citeauthoryear{{Kennedy}}{{Kennedy}}{2020}]{Kennedy20}
{Kennedy} G.~M.,  2020, \mn@doi [Royal Society Open Science]
  {10.1098/rsos.200063}, \href
  {https://ui.adsabs.harvard.edu/abs/2020RSOS....700063K} {7, 200063}

\bibitem[\protect\citeauthoryear{{Kennedy}, {Marino}, {Matr{\`a}}, {Pani{\'c}},
  {Wilner}, {Wyatt}  \& {Yelverton}}{{Kennedy} et~al.}{2018}]{Kennedy18}
{Kennedy} G.~M.,  {Marino} S.,  {Matr{\`a}} L.,  {Pani{\'c}} O.,  {Wilner} D.,
  {Wyatt} M.~C.,   {Yelverton} B.,  2018, \mn@doi [\mnras]
  {10.1093/mnras/sty135}, \href
  {https://ui.adsabs.harvard.edu/abs/2018MNRAS.475.4924K} {475, 4924}

\bibitem[\protect\citeauthoryear{{Lovell} \& {Lynch}}{{Lovell} \&
  {Lynch}}{2023}]{Lovell23}
{Lovell} J.~B.,  {Lynch} E.~M.,  2023, \mn@doi [\mnras]
  {10.1093/mnrasl/slad083}, \href
  {https://ui.adsabs.harvard.edu/abs/2023MNRAS.tmpL..83L} {}

\bibitem[\protect\citeauthoryear{{Lovell} et~al.,}{{Lovell}
  et~al.}{2021}]{Lovell21}
{Lovell} J.~B.,  et~al., 2021, \mn@doi [\mnras] {10.1093/mnras/stab1678}, \href
  {https://ui.adsabs.harvard.edu/abs/2021MNRAS.506.1978L} {506, 1978}

\bibitem[\protect\citeauthoryear{{Lovell} et~al.,}{{Lovell}
  et~al.}{2022}]{Lovell22}
{Lovell} J.~B.,  et~al., 2022, \mn@doi [\mnras] {10.1093/mnras/stac2782}, \href
  {https://ui.adsabs.harvard.edu/abs/2022MNRAS.517.2546L} {517, 2546}

\bibitem[\protect\citeauthoryear{{Lynch} \& {Lovell}}{{Lynch} \&
  {Lovell}}{2022}]{Lynch22}
{Lynch} E.~M.,  {Lovell} J.~B.,  2022, \mn@doi [\mnras]
  {10.1093/mnras/stab3566}, \href
  {https://ui.adsabs.harvard.edu/abs/2022MNRAS.510.2538L} {510, 2538}

\bibitem[\protect\citeauthoryear{{MacGregor} et~al.,}{{MacGregor}
  et~al.}{2017}]{MacGregor17}
{MacGregor} M.~A.,  et~al., 2017, \mn@doi [\apj] {10.3847/1538-4357/aa71ae},
  \href {https://ui.adsabs.harvard.edu/abs/2017ApJ...842....8M} {842, 8}

\bibitem[\protect\citeauthoryear{{MacGregor} et~al.,}{{MacGregor}
  et~al.}{2022}]{MacGregor22}
{MacGregor} M.~A.,  et~al., 2022, \mn@doi [\apjl] {10.3847/2041-8213/ac7729},
  \href {https://ui.adsabs.harvard.edu/abs/2022ApJ...933L...1M} {933, L1}

\bibitem[\protect\citeauthoryear{{Marino}, {Yelverton}, {Booth}, {Faramaz},
  {Kennedy}, {Matr{\`a}}  \& {Wyatt}}{{Marino} et~al.}{2019}]{Marino19}
{Marino} S.,  {Yelverton} B.,  {Booth} M.,  {Faramaz} V.,  {Kennedy} G.~M.,
  {Matr{\`a}} L.,   {Wyatt} M.~C.,  2019, \mn@doi [\mnras]
  {10.1093/mnras/stz049}, \href
  {https://ui.adsabs.harvard.edu/abs/2019MNRAS.484.1257M} {484, 1257}

\bibitem[\protect\citeauthoryear{{Matr{\`a}}, {Wyatt}, {Wilner}, {Dent},
  {Marino}, {Kennedy}  \& {Milli}}{{Matr{\`a}} et~al.}{2019}]{Matra19}
{Matr{\`a}} L.,  {Wyatt} M.~C.,  {Wilner} D.~J.,  {Dent} W.~R.~F.,  {Marino}
  S.,  {Kennedy} G.~M.,   {Milli} J.,  2019, \mn@doi [\aj]
  {10.3847/1538-3881/ab06c0}, \href
  {https://ui.adsabs.harvard.edu/abs/2019AJ....157..135M} {157, 135}

\bibitem[\protect\citeauthoryear{{Matthews}, {Krivov}, {Wyatt}, {Bryden}  \&
  {Eiroa}}{{Matthews} et~al.}{2014}]{Matthews14}
{Matthews} B.~C.,  {Krivov} A.~V.,  {Wyatt} M.~C.,  {Bryden} G.,   {Eiroa} C.,
  2014, in {Beuther} H.,  {Klessen} R.~S.,  {Dullemond} C.~P.,   {Henning} T.,
  eds, Protostars and Planets VI. pp 521--544 (\mn@eprint {arXiv} {1401.0743}),
  \mn@doi{10.2458/azu_uapress_9780816531240-ch023}

\bibitem[\protect\citeauthoryear{{Miller}}{{Miller}}{1971}]{Miller71}
{Miller} R.~H.,  1971, \mn@doi [\apss] {10.1007/BF00649196}, \href
  {https://ui.adsabs.harvard.edu/abs/1971Ap&SS..14...73M} {14, 73}

\bibitem[\protect\citeauthoryear{{Ogilvie}}{{Ogilvie}}{2001}]{Ogilvie01}
{Ogilvie} G.~I.,  2001, \mn@doi [\mnras] {10.1046/j.1365-8711.2001.04416.x},
  \href {http://adsabs.harvard.edu/abs/2001MNRAS.325..231O} {325, 231}

\bibitem[\protect\citeauthoryear{{Ogilvie} \& {Barker}}{{Ogilvie} \&
  {Barker}}{2014}]{Ogilvie14}
{Ogilvie} G.~I.,  {Barker} A.~J.,  2014, \mn@doi [\mnras]
  {10.1093/mnras/stu1795}, \href
  {http://adsabs.harvard.edu/abs/2014MNRAS.445.2621O} {445, 2621}

\bibitem[\protect\citeauthoryear{{Ogilvie} \& {Lynch}}{{Ogilvie} \&
  {Lynch}}{2019}]{Ogilvie19}
{Ogilvie} G.~I.,  {Lynch} E.~M.,  2019, \mn@doi [\mnras]
  {10.1093/mnras/sty3436}, \href
  {https://ui.adsabs.harvard.edu/abs/2019MNRAS.483.4453O} {483, 4453}

\bibitem[\protect\citeauthoryear{{Olofsson} et~al.,}{{Olofsson}
  et~al.}{2019}]{Olofsson19}
{Olofsson} J.,  et~al., 2019, \mn@doi [\aap] {10.1051/0004-6361/201935998},
  \href {https://ui.adsabs.harvard.edu/abs/2019A&A...630A.142O} {630, A142}

\bibitem[\protect\citeauthoryear{{Pan}, {Nesvold}  \& {Kuchner}}{{Pan}
  et~al.}{2016}]{Pan16}
{Pan} M.,  {Nesvold} E.~R.,   {Kuchner} M.~J.,  2016, \mn@doi [\apj]
  {10.3847/0004-637X/832/1/81}, \href
  {https://ui.adsabs.harvard.edu/abs/2016ApJ...832...81P} {832, 81}

\bibitem[\protect\citeauthoryear{{Peiris} \& {Tremaine}}{{Peiris} \&
  {Tremaine}}{2003}]{Peiris03}
{Peiris} H.~V.,  {Tremaine} S.,  2003, \mn@doi [\apj] {10.1086/378638}, \href
  {https://ui.adsabs.harvard.edu/abs/2003ApJ...599..237P} {599, 237}

\bibitem[\protect\citeauthoryear{{Rafikov}}{{Rafikov}}{2023}]{Rafikov23}
{Rafikov} R.~R.,  2023, \mn@doi [\mnras] {10.1093/mnras/stac3411}, \href
  {https://ui.adsabs.harvard.edu/abs/2023MNRAS.519.5607R} {519, 5607}

\bibitem[\protect\citeauthoryear{Schneider et~al.,}{Schneider
  et~al.}{2018}]{Schneider18}
Schneider G.,  et~al., 2018, \mn@doi [The Astronomical Journal]
  {10.3847/1538-3881/aaa3f3}, 155, 77

\bibitem[\protect\citeauthoryear{{Sefilian} \& {Rafikov}}{{Sefilian} \&
  {Rafikov}}{2019}]{Sefilian19}
{Sefilian} A.~A.,  {Rafikov} R.~R.,  2019, \mn@doi [\mnras]
  {10.1093/mnras/stz2412}, \href
  {https://ui.adsabs.harvard.edu/abs/2019MNRAS.489.4176S} {489, 4176}

\bibitem[\protect\citeauthoryear{{Sefilian}, {Rafikov}  \& {Wyatt}}{{Sefilian}
  et~al.}{2021}]{Sefilian21}
{Sefilian} A.~A.,  {Rafikov} R.~R.,   {Wyatt} M.~C.,  2021, \mn@doi [\apj]
  {10.3847/1538-4357/abda46}, \href
  {https://ui.adsabs.harvard.edu/abs/2021ApJ...910...13S} {910, 13}

\bibitem[\protect\citeauthoryear{{Sefilian}, {Rafikov}  \& {Wyatt}}{{Sefilian}
  et~al.}{2023}]{Sefilian23}
{Sefilian} A.~A.,  {Rafikov} R.~R.,   {Wyatt} M.~C.,  2023, \mn@doi [arXiv
  e-prints] {10.48550/arXiv.2305.00951}, \href
  {https://ui.adsabs.harvard.edu/abs/2023arXiv230500951S} {p. arXiv:2305.00951}

\bibitem[\protect\citeauthoryear{{Sende} \& {L{\"o}hne}}{{Sende} \&
  {L{\"o}hne}}{2019}]{Sende19}
{Sende} J.~A.,  {L{\"o}hne} T.,  2019, \mn@doi [\aap]
  {10.1051/0004-6361/201935199}, \href
  {https://ui.adsabs.harvard.edu/abs/2019A&A...631A.141S} {631, A141}

\bibitem[\protect\citeauthoryear{{Stark}, {Ren}, {MacGregor}, {Howard}, {Hurt},
  {Weinberger}, {Schneider}  \& {Choquet}}{{Stark} et~al.}{2023}]{Stark23}
{Stark} C.~C.,  {Ren} B.,  {MacGregor} M.~A.,  {Howard} W.~S.,  {Hurt} S.~A.,
  {Weinberger} A.~J.,  {Schneider} G.,   {Choquet} E.,  2023, \mn@doi [\apj]
  {10.3847/1538-4357/acbb64}, \href
  {https://ui.adsabs.harvard.edu/abs/2023ApJ...945..131S} {945, 131}

\bibitem[\protect\citeauthoryear{{Statler}}{{Statler}}{1999}]{Statler99}
{Statler} T.~S.,  1999, \mn@doi [\apjl] {10.1086/312306}, \href
  {https://ui.adsabs.harvard.edu/abs/1999ApJ...524L..87S} {524, L87}

\bibitem[\protect\citeauthoryear{{Statler}, {King}, {Crane}  \&
  {Jedrzejewski}}{{Statler} et~al.}{1999}]{Statler99a}
{Statler} T.~S.,  {King} I.~R.,  {Crane} P.,   {Jedrzejewski} R.~I.,  1999,
  \mn@doi [\aj] {10.1086/300737}, \href
  {https://ui.adsabs.harvard.edu/abs/1999AJ....117..894S} {117, 894}

\bibitem[\protect\citeauthoryear{{Touma}}{{Touma}}{2002}]{Touma02}
{Touma} J.~R.,  2002, \mn@doi [\mnras] {10.1046/j.1365-8711.2002.05437.x},
  \href {https://ui.adsabs.harvard.edu/abs/2002MNRAS.333..583T} {333, 583}

\bibitem[\protect\citeauthoryear{{Touma} \& {Sridhar}}{{Touma} \&
  {Sridhar}}{2012}]{Touma12}
{Touma} J.~R.,  {Sridhar} S.,  2012, \mn@doi [\mnras]
  {10.1111/j.1365-2966.2012.21000.x}, \href
  {https://ui.adsabs.harvard.edu/abs/2012MNRAS.423.2083T} {423, 2083}

\bibitem[\protect\citeauthoryear{{Touma} \& {Tremaine}}{{Touma} \&
  {Tremaine}}{2014}]{Touma14}
{Touma} J.,  {Tremaine} S.,  2014, \mn@doi [arXiv e-prints]
  {10.48550/arXiv.1401.5534}, \href
  {https://ui.adsabs.harvard.edu/abs/2014arXiv1401.5534T} {p. arXiv:1401.5534}

\bibitem[\protect\citeauthoryear{{Tremaine}}{{Tremaine}}{1995}]{Tremaine95}
{Tremaine} S.,  1995, \mn@doi [\aj] {10.1086/117548}, \href
  {https://ui.adsabs.harvard.edu/abs/1995AJ....110..628T} {110, 628}

\bibitem[\protect\citeauthoryear{{Tremaine}}{{Tremaine}}{2001}]{Tremaine01}
{Tremaine} S.,  2001, \mn@doi [\aj] {10.1086/319398}, \href
  {http://adsabs.harvard.edu/abs/2001AJ....121.1776T} {121, 1776}

\bibitem[\protect\citeauthoryear{{Wyatt}}{{Wyatt}}{2008}]{Wyatt08}
{Wyatt} M.~C.,  2008, \mn@doi [\araa] {10.1146/annurev.astro.45.051806.110525},
  \href {https://ui.adsabs.harvard.edu/abs/2008ARA&A..46..339W} {46, 339}

\bibitem[\protect\citeauthoryear{{Wyatt}, {Dermott}, {Telesco}, {Fisher},
  {Grogan}, {Holmes}  \& {Pi{\~n}a}}{{Wyatt} et~al.}{1999}]{Wyatt99}
{Wyatt} M.~C.,  {Dermott} S.~F.,  {Telesco} C.~M.,  {Fisher} R.~S.,  {Grogan}
  K.,  {Holmes} E.~K.,   {Pi{\~n}a} R.~K.,  1999, \mn@doi [\apj]
  {10.1086/308093}, \href {http://adsabs.harvard.edu/abs/1999ApJ...527..918W}
  {527, 918}

\makeatother
\end{thebibliography}







\bsp	
\label{lastpage}
\end{document}